# SOFT X-RAY REFLECTIVITY: FROM QUASI-PERFECT MIRRORS TO ACCELERATOR WALLS

F. Schäfers, Institute for Nanometre Optics and Technology, HZB BESSY-II, Berlin, Germany

R. Cimino, LNF - INFN, Frascati, Italy

*Abstract*

Reflection of light from surfaces is a very common, but complex phenomenon not only in science and technology, but in every day life. The underlying basic optical principles have been developed within the last five centuries using visible light available from the sun or other laboratory light sources. X-rays were detected in 1895, and the full potential of soft- and hard-x ray radiation as a probe for the electronic and geometric properties of matter, for material analysis and its characterisation is available only since the advent of synchrotron radiation sources some 50 years ago. On the other hand high-brilliance and high power synchrotron radiation of present-days $3^{rd}$ and $4^{th}$ generation light sources is not always beneficial. High-energy machines and accelerator-based light sources can suffer from a serious performance drop or limitations due to interaction of the synchrotron radiation with the accelerator walls, thus producing clouds of photoelectrons (e-cloud) which in turn interact with the accelerated beam. Thus the suitable choice of accelerator materials and their surface coating, which determines the x-ray optical behaviour is of utmost importance to achieve ultimate emittance. Basic optical principles and examples on reflectivity for selected materials are given here.

## INTRODUCTION

Basic optical principles of light reflection have been developed within the last five centuries by the pioneering work of Snell, Huygens, Brewster, Fresnel, Maxwell and many others using visible light arising mainly from the sun. X-rays were detected in 1895 by Wilhelm Conrad Röntgen, and in the years to follow it was realised that they are part of the electromagnetic spectrum and can be treated with the same mathematical and optical formalism. For modern reviews about optics see ref. [1-5]. The full potential of soft- and hard-x ray radiation for material analysis and characterisation is available since the advent of synchrotron radiation (SR) sources some 50 years ago. After the first parasitic use of SR at high-energy accelerators the second generation facilities were explicitly dedicated to the production and use of SR from bending magnets. Since 20 years the $3^{rd}$ generation storage ring facilities make use of high-intense and high-brilliance undulator radiation. The upcoming $4^{th}$ generation light sources are Free Electron Lasers (FEL) extending the SASE-lasing principle up to the hard x-ray range. They provide full coherent femto-second radiation pulses with a peak-brilliance up to $10^{33}$ Ph/s/mm$^2$/mrad$^2$ and peak power densities up to $10^{20}$ Watt/cm$^2$/shot [6].

For many researchers this radiation is an ideal probe of the electronic or geometric properties of matter, for machine operators it is an unavoidable problem, which prevents the ultimate performance of the machine to be achieved. High-energy collider machines [7-11] and accelerator-based light sources with positive beams [12, 13] suffer from serious performance drop or limitations due to the interaction of the light with the accelerator walls, by which clouds of photoelectrons are created. Thus the material choice, its photoelectric and x-ray optical behaviour, is of utmost importance for ultimate performance accelerators [14, 15, 16].

In this contribution basic principles of the reflection on boundaries and examples for the optical characterisation of materials are given. The key parameter to determine the optical properties for a material is the shape, the waviness and roughness of its surface. The instrumentation for the optical characterisation of a surface from atomic level to long spatial frequencies is described. The main characterisation tool is at-wavelength metrology, the measurement of the reflectivity of a material at the wavelength and incidence angle of interest, which is regarded as the test drive for optical elements.

## LOW ENERGY ELECTRONS IN ACCELERATORS

Low-energy background electrons form the so called "electron cloud" in particle accelerators with positive beams, and can interfere with the correct operation of the machine itself; there are running conditions that can lead to large amplification of such electron clouds [7-11]. If the cloud density becomes sufficiently large, the beam/electron-cloud interaction can degrade the particle beam. In positively charged particle rings, such electron cloud can oscillate synchronously with the particle bunches, giving rise to an exponential growth of the electron density. In such a resonant phenomenon, electrons produced directly by ionization of the residual gas molecules or by irradiation of the vacuum chamber surfaces by synchrotron radiation, ions, beam particles, or electrons themselves, "see" the positive beam and, at each passage, gain energy. The beam accelerated electron cloud hit the accelerators walls, thus producing a cascade of low energy electrons according to the Secondary Electron Yield (SEY) value [17] (i.e. the number of electrons produced by an impinging electron) of the accelerator wall. The SEY value is a crucial parameter in accelerator technology and can vary, as studied widely in literature [18-22] according to surface quality and composition, ranging from less than one to, in some cases, more than three. The phenomenon just described is called *multipacting*, and may induce an anomalous vacuum pressure rise and it can cause beam instabilities. In order to predict, and possibly prevent these problems, some simulations must be performed on the formation and development of the electron cloud in accelerator rings. Obviously, photoelectrons, if present, are essential ingredients to determine when and were such a resonant phenomenon can take place, hence photo–reflectivity, photo-emission and electron-induced secondary electron production are to be considered essential ingredients to be used in the simulation codes [14, 22].

The importance of studying the reflectivity and photo yield is clear when considering what may occur in the arcs of an accelerator. The produced synchrotron radiation, with a given beam divergence, will illuminate the accelerator walls at a grazing incidence angle. Most of the photon beam will be scattered/reflected away, some will create photoelectrons in the presence of the dipole magnetic field perpendicular to the orbit plane. The electrons photoemitted in the orbit plane, being affected by the magnetic field, are constrained to move along the field lines, thus they will not be able to cross the vacuum chamber and gain energy from the beam. On the other hand, the scattered photon beam will soon illuminate top and bottom walls, emitting photoelectrons perpendicular to the orbital plane (hence parallel to the magnetic field) that will only spiral along the field lines, efficiently participating in secondary electron production and, eventually, in multipacting. This simple reasoning shows how important it is to determine the photon reflectivity of accelerator walls and its photo yield experimentally [15, 16, 22].

E-cloud related resonant phenomena are not the only detrimental effects occurring in accelerators and originated by the background electrons. Ohmi and Zimmermann [23], in 2000, when attempting to explain the observed vertical beam-size blow up for KEKB, introduced the concept of "single beam instability threshold" suggesting that the mere existence of a certain electron density in the accelerator (for the KEKB case around $7e^{11}$ e-/m$^3$) is able to detrimentally affect the beam quality. Hence, even in the absence of resonant phenomena, such electron density has to be carefully simulated, controlled and careful material choice is needed for its mitigation. This paper presents the tools used in other field of physics for similar problems and analyse their potential and outreach in technology oriented accelerator science.

## SYNCHROTRON RADIATION

The properties of synchrotron radiation (SR) as well as its applications for various disciplines of science and material research are described in many lecture books [see, for instance, 24-26]. Charged particles moving close to the speed of light when deflected by magnets irradiate a very special light 'in all colours' with unique properties: It covers the spectral range from infrared up to the x-ray range. The high energy limit is given by the electron energy. SR is highly collimated (< 1 mrad) perpendicular to the plane of the orbit with high intensity. It is polarised linearly within- and elliptically outside the orbit plane. It is a pulsed beam with a well defined time structure of pulse widths in the picosecond range and repetition rates in the nanosecond range. All properties of SR can be easily calculated.

Typical bending magnet beamlines have an acceptance angle of some milliradians only, that is, at most SR-facilities not more than approximately 1 % of the total radiation is transported into SR-beamlines and to user experiments. 99 % of the emitted power which may be more than some tens of kW is left inside the accelerator tube.

Third generation storage rings primarily make use of undulator radiation. Periodic magnetic devices, built into the straight sections of the storage ring create quasi-monochromatic radiation with very high peak brilliances in the first and higher harmonics. The photon energy is tuneable by varying the magnetic field on-axis, i.e. at permanent magnet undulator designs, the undulator gap. Helical undulators (e.g. APPLE II design) provide a polarisation which is tuneable between linear and circular by varying the undulator shift - the lateral translation between the top and bottom magnets.

## PHOTON MATTER INTERACTION

An accelerator tube is filled with radiation. It is brightly illuminated with a large portion of the electromagnetic spectrum. And this soup of radiation interacts with rest gas atoms and with the vacuum tube.

One of the key processes in nature is photon - matter interaction and this takes place in three ways:

- Photoelectric effect – radiative with emission of another photon or non-radiative with emission of a photoelectron;
- Scattering – either elastic (photon preserves energy) or inelastic (photon loses energy);
- Pair production with generation of an electron - positron pair. The minimum energy for this process is 0.5 MeV. So it is negligible in the SR range.

The photoelectric effect is the most dominant process in the x-ray range up to some hundred keV. So this one will be treated here only.

A brief summary of text book optics is given now which has been compiled centuries ago by the pioneering work of Willebrord von Ruijen Snell (1580 – 1626), Christiaan Huygens (1629 - 1695), David Brewster (1781 - 1868), Augustin Fresnel (1788 - 1827), James

Clark Maxwell (1831 - 1879), and many other researchers making experiments with visible light.

An electromagnetic wave is described in a suitable coordinate system (*s*: senkrecht (= perpendicular); *p*: parallel to the polarisation plane) by its amplitude $E_{s,p}$ and its phase $\phi_{s,p}$:

$$E_{s,p}(r,t) = E_{s,p}^0 e^{(i\omega t - kr - \phi_{s,p})} \quad (1)$$

with the intensity:

$$I_0 = |E_0|^2 \quad (2)$$

At the boundary of two media described by the optical constants $n_1$ and $n_2$ an interaction takes place. The incident wave is split into a reflected and a refracted wave. This is described by the complex reflection and transmission coefficients:

$$r_{s,p} = |r|e^{i\delta_{s,p}} \quad (3)$$

and

$$t_{s,p} = |t|e^{i\delta_{s,p}} \quad (4)$$

with the reflected and transmitted intensities:

$$R_{s,p} = |r_{s,p}|^2 \quad \text{and} \quad T_{s,p} = |t_{s,p}|^2 \quad (5,6)$$

The Fresnel equations which are derived from Maxwell's equations with adequate boundary conditions determine the reflection and transmission coefficients on the basis of the incidence and reflection angle $\theta_i$, the refraction angle $\theta_t$ and the optical constants of the two media $n_1$ and $n_2$:

$$r_s = \frac{E_{rs}}{E_{is}} = \frac{n_1 \cdot \cos\theta_i - n_2 \cdot \cos\theta_t}{n_1 \cdot \cos\theta_i + n_2 \cdot \cos\theta_t}$$

$$t_s = \frac{E_{ts}}{E_{is}} = \frac{2n_1 \cdot \cos\theta_i}{n_1 \cdot \cos\theta_i + n_2 \cdot \cos\theta_t}$$

$$r_p = \frac{E_{rp}}{E_{ip}} = \frac{n_2 \cdot \cos\theta_i - n_1 \cdot \cos\theta_t}{n_1 \cdot \cos\theta_t + n_2 \cdot \cos\theta_i}$$

$$t_p = \frac{E_{tp}}{E_{ip}} = \frac{2n_1 \cdot \cos\theta_i}{n_1 \cdot \cos\theta_t + n_2 \cdot \cos\theta_i} \quad (7)$$

The interaction is governed by one number *n* - namely the complex index of refraction or the optical constant

$$n(\lambda) = 1 - \delta - ik \quad (8)$$

with the real part $\delta$ describing the reflection and $k$ determining the absorption properties of a material. This number is wavelength dependent and it is compiled practically for the whole electromagnetic range [27-30].

Figure 1 lists a number of optical data tables available for the optical constants from the visible to the x-ray range which are incorporated in open access reflectivity programs such as REFLEC [31].

For selected molecules there are experimental molecular data bases, if this is not the case, in REFLEC the molecule is decomposed into atoms and the optical constants are stochiometrically composed from the individual atomic data compiled by Palik [27] (UV, soft x-ray), Henke [28, 29] (30 eV – 30 keV) or by Cromer [30] (5 – 100 keV) .

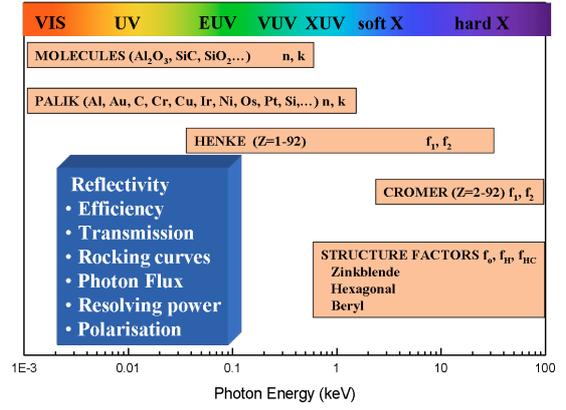

Figure 1: Optical data tables incorporated in the REFLEC-program [31]

To calculate crystal diffraction the structure factors for the special cases of Zinkblende and Hexagonal crystals are calculated by REFLEC.

These optics library tools make calculations of the transmission and reflectivity of any material or (multilayer-) coating as function of the wavelength or the incidence angle, calculation of grating efficiencies or crystal rocking curves, and polarisation effects possible.

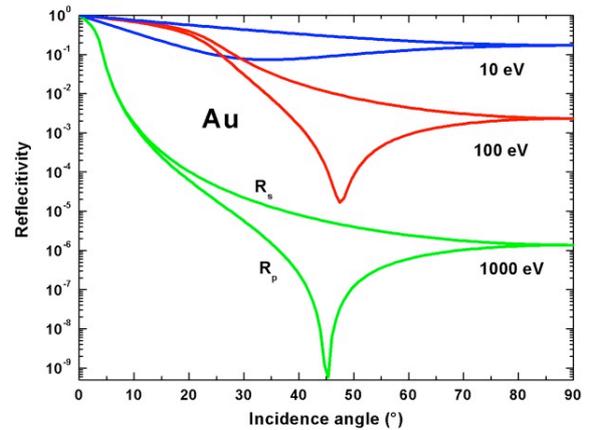

Figure 2: Reflectivity $R_s$ and $R_p$ of a perfect Au-coating as function of incidence angle in the UV and soft x-ray range.

Figure 2 shows as an example the reflectivity calculated by the Fresnel equations for an Au-surface at three photon energies of 10 eV, 100 eV and 1000 eV, plotted for both reflectivity components $R_s$ and $R_p$. At small angles, in the total external reflection regime, all light is reflected. Note that $R_s$ decreases gradually, while $R_p$ goes through a Brewster minimum, which is at around 45° in the x-ray range ($\theta_B=atan(1/n)$).

Significant normal incidence reflectivity ($\theta=90°$) is available only in the visible and the UV spectral range. This is the case for all materials. Therefore mirrors with significant reflectivity above 30 eV have to be operated under grazing incidence angles. The polarisation of the light is not significantly altered at grazing and normal incidence angles ($R_s=R_p$), while depolarisation effects occur at intermediate angles. This is exploited in polarisation steering and monitoring devices on the basis of reflection optical elements which operate close to the Brewster-angle to achieve maximum polarization ($R_s>>R_p$) [32, 33].

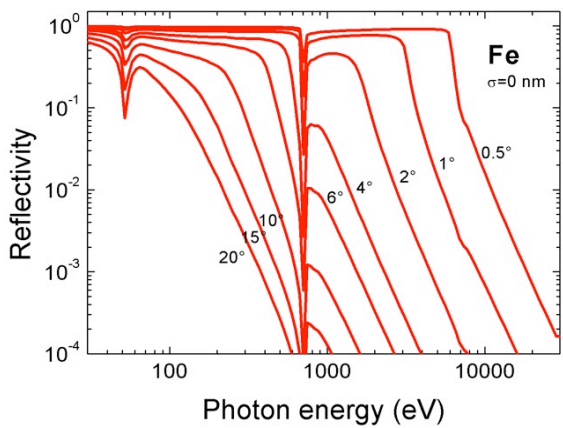

Figure 3: Fresnel-reflectivity of a smooth Fe surface as function of photon energy for various incidence angles.

More realistic materials to be used as accelerator walls are Cu, Al or Stainless Steel (SS). The Fresnel reflectivity of e.g. pure supersmooth Fe is plotted in Fig. 3 as function of the photon energy for selected incidence angles between 0.5° and 20°.

Reflectivity is always decreasing with angle. The angle determines the high energy cut-off. At the Fe 3p (52.7 eV) and 2p (708 eV) absorption edges the reflectivity goes down since absorption increases.

## FROM MIRRORS TO ACCELERATOR WALLS

So far only perfect interfaces - perfect optical coatings with atomic scale smoothness have been discussed. Fresnel-equations do not know anything on real surfaces or even accelerator tubes.

To take this into account, consider the surface to be composed of individual facets, with each surface normal having an individual angle $\Delta\alpha_i$ with respect to the mean surface plane. Initially parallel incident beams will then be deflected by $2\cdot\Delta\alpha_i$. The rms-value of the angular distribution of the surface facets $\Delta\alpha_i$, assuming a Gaussian probability distribution of the facets orientation, is usually called the slope error of the surface (Fig. 4). The slope errors gives rise to a blurring of the reflected beam, to straylight and thus in turn to a reduction of the specular (Fresnel-) reflectivity $R_o$ which is taken into account by a Debye-Waller factor according to:

$$R = R_0 e^{-\left(\frac{4\pi\sigma \sin\theta_i}{\lambda}\right)^2} \qquad (9)$$

the surface roughness σ can be interpreted as the rms-value of the height deviation, λ is the wavelength. Note that FWHM=2.3σ. With this statistical concept of slope $\Delta\alpha$ and roughness σ, which can be transformed into each other taking into account the frequency, (see below), and which are directly accessible from optical metrology, the specification of surface figure and finish of X-ray mirrors in terms of their imaging performance can be addressed [34, 35].

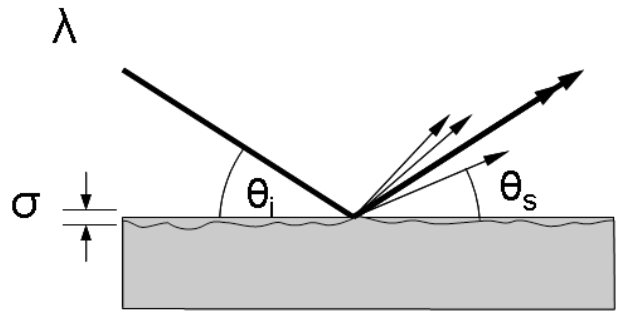

Figure 4: Surface roughness σ leads to a blurring of the reflected beam.

The dramatic influence of the roughness σ onto the specular reflectivity is illustrated in Fig. 5 for the case of e.g. a Cu sample and calculated for an x-ray wavelength of 1 nm.

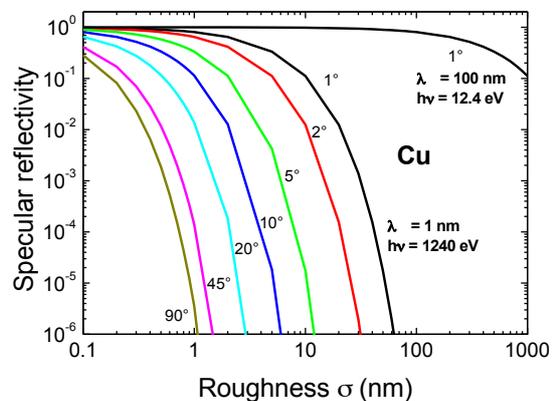

Figure 5: Specular (Fresnel) reflectivity of Cu as function of its surface roughness for an x-ray wavelength of 1 nm and 100 nm (1° curve only) (calculation with REFLEC).

Obviously high-quality x-ray mirrors require a roughness in the sub-nanometer range. Machined surfaces with roughness values up to 1 μm have no noticeable x-ray specular reflectivity and produce nothing but an isotropically scattered light distribution. For a statistical quantitative description of a surface profile consisting of individual facets of different angles, and in order to introduce the frequency of slope and roughness, the concept of the Power Spectral Density PSD is introduced. It is plotted double-logarithmically as function of the tangential frequency, which is the 'grating' period of the facets. An example is shown in Fig. 6. Three ranges can be identified: a low (LSF), mid (MSF) and a high spatial frequency (HSF) range. The impact on the optical properties depends strongly on the frequency. Long waves produce a wavefront distortion, i.e. a phase mismatch on the reflected beam, while mid frequencies are responsible for small angle scatter. HSF results in wide angle scatter. Obviously the concepts of slope error of a surface (which corresponds to an angular mismatch) and roughness (which corresponds to a height distribution) can be regarded as one and the same. It's just a different approach according to the frequency range, which is sampled experimentally [36].

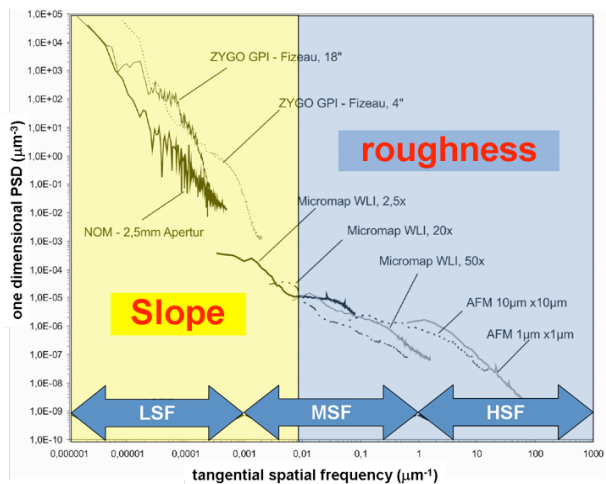

Figure 6: Power Spectral density curve of an x-ray mirror [36].

The PSD covers more than 10 orders of magnitude in frequency range and in magnitude. It is mapped experimentally by various methods, such as interferometry (ZYGO) and profilometry (LTP, NOM) [37, 38], to detect the low frequencies, while Atomic Force Microscopy scans the high frequencies. Finally, x-ray scattering samples the highest spatial frequencies according to the wavelength used. X-ray mirror quality surfaces have a smooth diagonal line in such a plot across the whole frequency range.

The optical behaviour of such a surface, seen as sum over gratings of different periods, can be comprised in the following equation, which gives the angular distribution of the scattered or reflected power [34, 35, 39]:

$$\frac{1}{P_0}\frac{dP}{d\omega} = \left(\frac{16\pi^2}{\lambda^4}\right) (R) \left(\sin^3\theta_i\right) \left(PSD_{2D,\,2\,sided}(\mathbf{f})\right) \quad (10)$$

This is the power $P$ distributed into a solid angle $d\omega$, normalised to the incident power $P_o$. It contains four terms:
- The Rayleigh blue sky factor: Scattering goes up very fast, with the fourth power of energy;
- The material properties are summarised in $R$;
- Scattering goes up with the third power of the incidence angle
- Power spectral density $PSD$ of a surface can be mapped experimentally.

This equation is valid in the 'smooth' surface roughness region only, i.e. polished surfaces. To our knowledge there is no mathematical formalism dedicated for machined, rough surfaces available. The optical behaviour of surfaces of any quality can, however, be simulated by intensity raytracing codes such as RAY [40], provided the real two-dimensional surface profile is available from experimental mapping.

State of the art x-ray optical elements necessary for nanofocussing synchrotron radiation beamlines have a 0.05 μrad (0.01″) slope error plane or focusing mirrors. This corresponds to a roughness of less than 1 nm rms. Mirrors for hard x-ray FEL radiation require the same slope, but on a length of 800 mm. The peak-to-valley (p-v) of such a mirror corresponds to 2 nm. This requires a challenging technology for both shaping the mirrors by advanced and ion beam polishing techniques and also characterising them with slope measuring devices such as the BESSY-II Nanometre Optics Measuring machine (NOM) [35-36].

## REFLECTOMETRY

Reflectometry or at-wavelength metrology is a powerful and most essential characterisation tool for the development and characterisation of optical elements [41-43]. With this method the reflectivity of a material (mirrors, crystals), the diffraction efficiency of gratings or the transmission of thin films is investigated at the design wavelength for the optical element. Since the optical constants of the coating materials involved are dependent on wavelength, information on reflectivity at a certain wavelength can be obtained only by this method and cannot be deduced from any other diagnostics technique. Thus this method is complementary to the above mentioned ex-situ profilometry methods, or to Cu-Kα diffractometry which delivers information on interfacial roughness and quality of reflecting layers. It is the decisive test-drive for optics before delivery and installation.

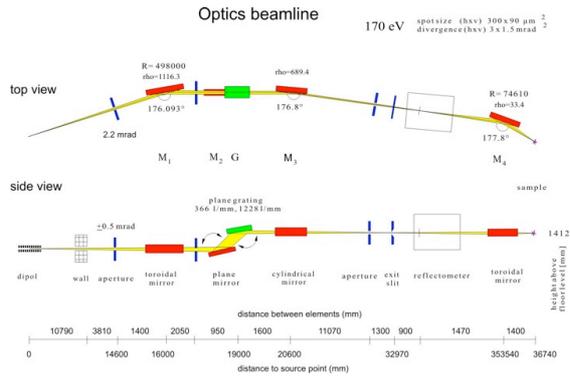

Figure 7: The soft x-ray optics beamline for at-wavelength metrology at BESSY-II [44]

At the Berlin BESSY-II storage ring facility an optics beamline has been dedicated for such at-wavelength metrology in the soft x-ray range. This plane grating monochromator (PGM) beamline is schematically shown in Fig. 7. It employs bending magnet radiation which is collimated vertically and focussed horizontally by a toroidal mirror, dispersed by a plane mirror / plane grating monochromator and refocused by a cylindrical mirror onto the exit slit with low divergence. Behind the exit slit a three-axis UHV-reflectometer chamber is attached permanently to the beamline. The operation of the PGM in collimated light [45] allows the monochromator to be optimised for either high flux, high energy resolution or for highest spectral purity (suppression of higher orders). By use of interchangeable gratings (1228, 366 l/mm) the available energies range is between 10 eV to 2 keV. The incident beam can be tuned to either linear or elliptical polarisation.

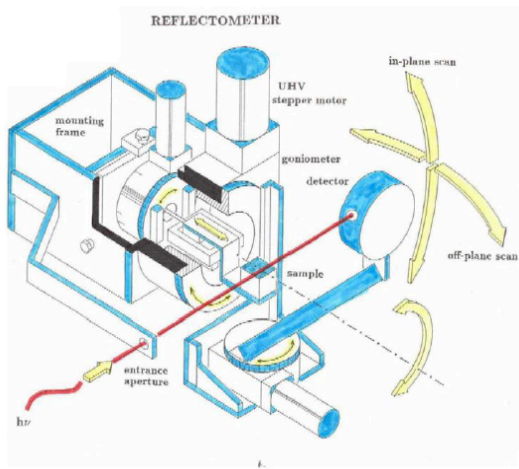

Figure 8: Schematic setup of the BESSY-II three-axis reflectometer with moveable sample and detector stage [46].

The reflectometer is schematically shown in Fig. 8. Two concentric HUBER-goniometers with UHV-motors rotate the sample and the detector stage around the incident beam axis between 0° and 90° (180°), and a third goniometer allows the detector to be scanned off-plane. The plane of reflection is vertical and since the light is horizontally polarised, the reflectivity is measured in s-polarisation geometry ($R_s$). Absolute values for $R_s$ are determined by a sample-in sample-out technique by measuring the incident intensity $I_o$ with the detector placed in the direct beam position before and after measuring the reflected intensity $I$ (to check stability). The absolute value of reflectivity is then given by $R_s = I/I_o$.

Standard measurement schemes with a reflectometer setup are shown in Fig. 9. Reflectivity is measured typically with a monochromatic beam as function of the incidence angle $\theta$, at a fixed photon energy $h\nu$, or as a function of the photon energy at a fixed incidence angle.

Scattered light is measured at fixed energy and incidence angle while varying only the detector angle $\theta_{det}$ around the specular beam. With these techniques information on surface roughness and interface quality and its chemical composition, (multi-)layer thickness, and, of course, optical constants can be addressed [47-48]. Coating thicknesses can be determined with sub-nanometer precision.

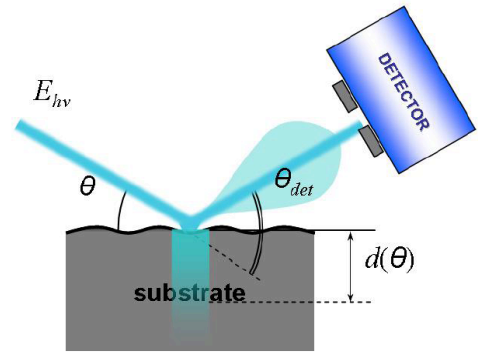

Figure 9: Geometry in reflectometry experiments.

In practise one must be aware of potential risks in a reflectometry experiment in order not to misinterpret the measured data:
- the sample's- rotation axis may not be exactly in the centre of the incident beam; and or
- the sample's surface may not be in the goniometer rotation axis;
- the footprint of the beam onto the sample may be longer than the sample at very grazing angles;
- the detector area must be larger than the beam cross section;
- sample and detector angle may be misadjusted;
- the sample may be tilted – and so the reflection plane, so that the detector moves out of the reflected beam at large angles.

At 0° incidence angle the Fresnel reflectivity is always 100 % (see Fig. 2), while the 'measurable' reflectivity will be 50 %, whenever the sample is perfectly adjusted. At 0° the surface of a perfectly aligned sample lies parallel to the beam axis intercepting half of it, and leaving the other half of the incoming beam free to reach the detector. In other words: a reflectivity cannot be measured at all at angles between $\theta=0°$ and $\theta < atan(h/L)$, (where: $h$=beam height, $L$=length of sample).

Since the sample has a finite length, there is a minimum angle larger than 0° at which experimental reflectivity curves start. Thus, the measured reflectivity curve especially at small angles, in the region of total reflection, is often dominated by alignment problems rather than yielding any information on the sample.

However, another information which is complementary to reflectivity data, is easily obtainable within the same experiment. This is the photo yield of the sample. This requires nothing but an electrically isolated sample, of which the photocurrent induced by the monochromatic synchrotron beam is simultaneously recorded. Having a calibrated light detector (e.g. a GaAsP-photodiode) [49] of which the efficiency as function of photon energy is known, this information can easily be converted into the number of photoelectrons per incident photon, once the number of incident photons/sec normalised to the ring current is known.

## EXPERIMENTAL RESULTS

To show the potentiality of the technique and of the available set-up, we present here some selected results without aiming to a comprehensive description of the whole data set collected. This will be done in forthcoming publications.

The data analysed here are on Cu technical samples representative of LHC inner beam screen [14 - 16].

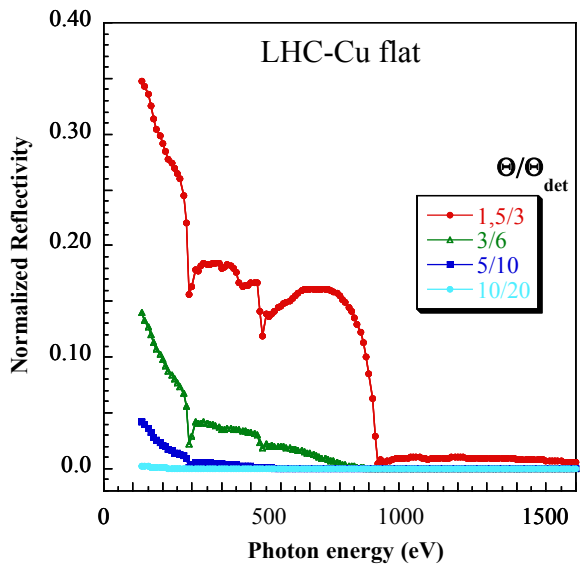

Figure 10: Reflectivity of LHC-Cu sample representative of the flat part of the beam screen, as function of photon energy for various incidence angles $\Theta$ and emission angle $\Theta_{det}$.

The samples, all about 100 mm long and 20 mm wide, were isolated from the sample holder by a kapton foil to allow the contemporary acquisition of reflectivity (as measured from the photodiode) and photo yield (as derived from the current emitted by the sample during irradiation with a well defined photon flux). In addition, for comparison purposes, a polished Si-plane mirror with a sub-nanometer roughness was also measured.

In Fig. 10 the reflectivity of such a LHC Cu sample is shown as function of the photon energy for selected incidence angles. The investigated energy range presented here is between 130 eV and 1600 eV. In Fig. 10 we observe higher reflectivity at lower grazing angle incidence and at low photon energy. We also observe resonance structures due to absorption of the Carbon and Oxygen contaminated surface layers (C, O k-edges at 284.4 eV and 543.1 eV) and of Cu (Cu $L_3$-edge at 932.7 eV). As expected, the reflectivity goes down with increasing incidence angle and energy.

In Fig. 11 we show the measured normalized reflectivity (in logarithmic scale) of an LHC Cu flat sample as a function of the emission angle $\Theta_{det}$ for a photon energy of $h\nu$=150 eV and a fixed incidence angle of $\Theta$.

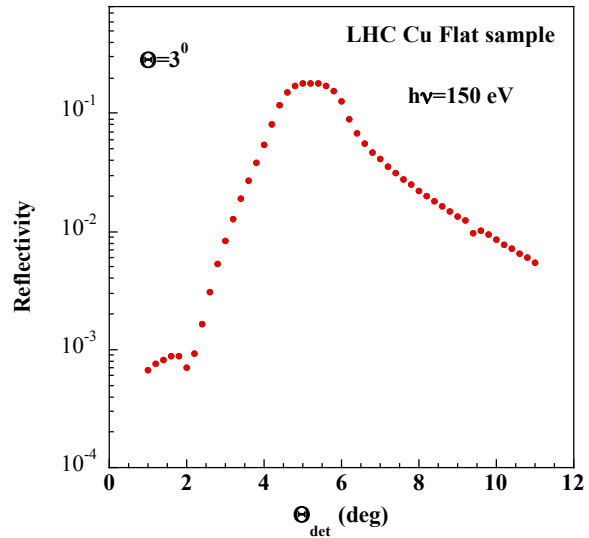

Figure 11: Normalized reflectivity of LHC- Cu Flat sample as function of emission angle $\Theta_{det}$, for a given photon energy $h\nu$=150 eV and incidence angle of $\Theta$=3°.

The maximum reflectivity is expected and observed at around 6°. In figure 11 we just see a broad angular continuum in the range of the $\Theta_{det}=2\theta$ peak with no clear separation of the specular reflection peak from scattered light. This is as expected, since this machined sample surface has mainly low-frequency roughness (long waves). This creates a blurred beam so that the detector, with an acceptance area of 4 mm x 4 mm, may be overfilled (see Fig. 4 and 5). This effect is so dominant that micro- roughness (i.e. wide angle scatter) cannot be identified here.

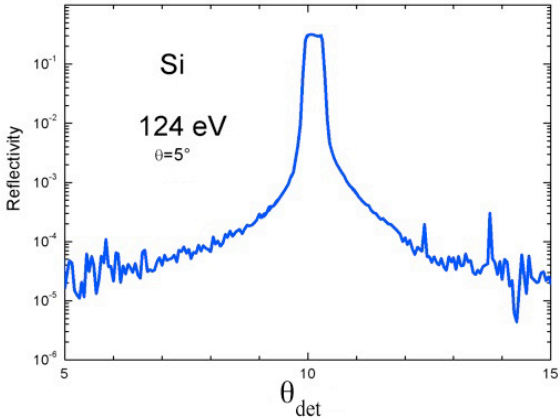

Figure 12: Scattering from a quasi-perfect Si mirror surface taken at 5° incidence angle and at 124 eV (10 nm). Data taken with a 4x4mm photodiode.

The comparison of this plot with Figs. 12 and 13 is enlightning. Fig. 12 shows scattering data ($\Theta_{det}$ scans) obtained from a quasi-perfect Si-mirror surface taken at θ=5° incidence angle at a photon energy of 124 eV (10 nm) with the 4x4mm photodiode, as in case of Fig 11. The high reflectivity plateau around 10° corresponds to $\Theta_{det}=2\theta$ positions where the small specularly reflected peak enters in the field of view of the unmasked photodiode, so it is essentially an artefact of the measurement. Such artificial plateau can be experimentally eliminated by masking the photodiode with a pinhole of a diameter similar to the beam dimension (300 μm) as done in Fig. 13. In any case, what is clear to occur on a quasi-perfect mirror is the extremely strong reduction of any reflectivity signal outside the specularly reflected beam.

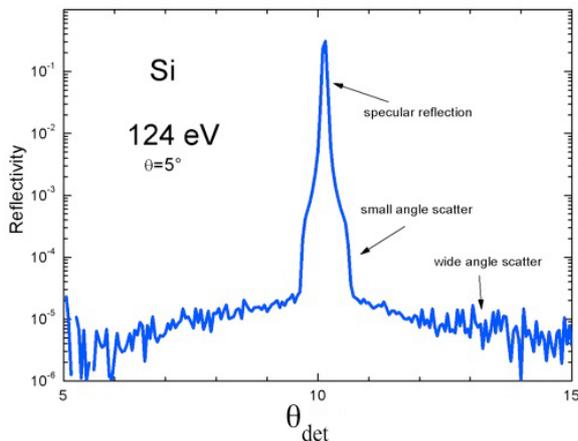

Figure 13: Scattering from a quasi-perfect Si mirror surface taken at 5° incidence angle and at 124 eV (10 nm). Data taken with a 4x4mm photodiode masked by a 0.25 mm pinhole

This is not the case for the measurement shown in Fig. 11 from the Cu technical surface. In Fig. 12 and 13 a quasi-perfect Si mirror shows a specularly reflected peak at $\Theta_{det}=2\theta=10°$ of 30 % reflectivity, connected with a symmetric scattering profile, which can be divided into a small angle scatter (MSF) and a wide angle scatter (HSF) region extending over the full solid angle. A Fourier-analysis of such a curve would yield the frequency spectrum (PSD) of the surface in this respective range of the sampling wavelength (10 nm). The measurable angular range is limited by the dynamic range of the photodiode of six orders of magnitude (Io-intensity: ~100 nA, dark current 30 fA). Single photon counting detectors would have to be applied to extend the curve to larger scattering angles. This is indeed an illustration of the power of soft x-ray reflectivity data for optical surfaces and its intrinsic difference with reflectivity from technical surfaces.

In Fig. 14 we show $\Theta/\Theta_{det}$ scans of LHC Cu flat sample at different photon energies. Such scans are obtained like the one in Figs. 11 and 12, but rather than keeping a fixed incidence angle $\Theta$ and just vary $\Theta_{det}$, we vary simultaneously the two angles, keeping $\Theta_{det}$ on the maximum of the specular reflection ($\Theta_{det}=2\theta$) for each incidence angle.

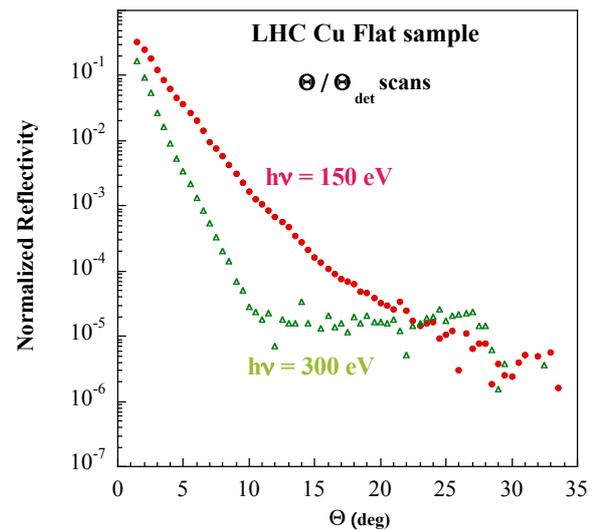

Figure 14: Normalized Reflectivity of LHC- Cu Flat sample as function of incidence angles $\Theta$ and emission angle $\Theta_{det}$ for two photon energies of 150 and 300 eV, respectively.

Figure 14 shows that, as expected, the specular reflectivity decreases with incidence angle and more significantly using higher energy photons. For accelerator studies on instabilities and e-cloud effects related to the presence of photoelectrons produced at accelerator walls, we show that photoelectrons need low grazing incidence reflections of low energy photons to be produced in places not directly illuminated by the primary photon beam.

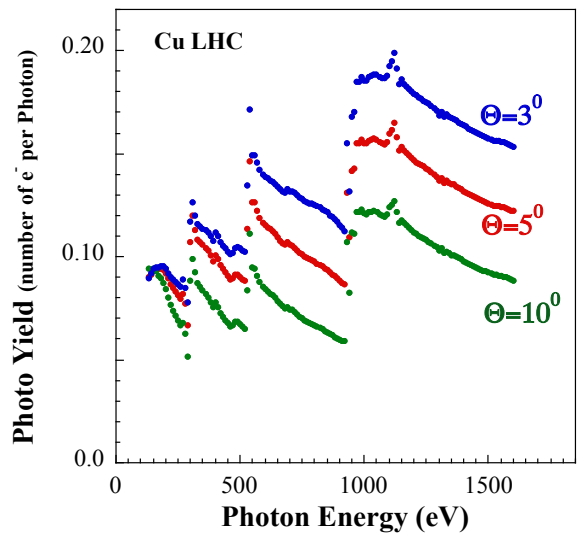

Figure 15: Photo yield (number of electrons emitted per incident photon) from a Cu technical surface of LHC beam screen, as function of photon energy at different incidence angles.

As previously discussed, not only photon reflectivity is an important ingredient for accelerator studies but also the efficiency of each sample to produce electrons once photo-irradiated. The complementary information we could measure during reflectivity runs is the photo yield of a sample. This is shown in Fig. 15 where the photo yield PY (number of electrons emitted per incident photon) of the LHC Cu sample (same as the one measured in Fig. 11 is plotted as function of the impinging photon energy for three different incident angles of $\Theta= 3°, 5°, 10°$. It was possible to extract PY data by normalizing the current measured from our samples by the number of photons impinging on them. We can estimate the number of incident photons as a function of their energy, by measuring the current produced when they directly impinge on the photodiode, and scale this current by the tabulated quantum efficiency of the calibrated light detector used (GaAsP from Hamamatsu).

As a general trend, we clearly observe the increase of the photon yield for increasing photon energy (which is expected considering that more energetic photons have more energy to donate to the low energy electrons which mainly contribute to the total PY) [14]. From Fig. 15, we also note that, for all photon energies, the PY increases at decreasing incident angles. This can be attributed to the reduced photon penetration at very grazing angles, so that, since more surface is irradiated at lower incidence angles, more electrons are produced by the photoelectric effect closer to the surface which are more ready to escape into vacuum.

Finally, we notice that using monochromatic light, the PY curves are essentially proportional to the sample absorption structure, which is not a smooth structure, but have singularities at absorption edges. Such edge structures are indeed observable in Fig. 15. Next to Cu L-absorption lines (Cu L-edge around 933 eV), the surface contaminating layer is observed. Both C and O are visible at their absorption edges (C, O k-edges at around 284 eV and 543 eV). Detailed analysis of such resonance regions around the absorption edges would yield further information on chemical composition of the coatings and interfaces, and on the individual thicknesses. For instance, it will be possible to characterize the chemical origin of the PY, its eventual modification by photon irradiation and compare it to what has been recently studied on electron irradiated Cu [20]. This is however, beyond the scope of our present investigation.

## CONCLUSIONS

We have shown that the optical behaviour of a material is strongly dependent on its surface properties and surface quality. Optical theory such as the Fresnel-reflectivity refers only to perfect atomic smooth surfaces which do not exist in reality. A description of a real polished, lapped or just machined surface needs to take the long waves as well as the (micro-) roughness into account, of which the frequency varies over more than 10 orders of magnitude. This can be covered only by statistical methods which leads to the Power Spectral Density (PSD) of a surface. A variety of surface characterisation tools such as profilometry, interferometry and AFM for measuring the PSD have to be applied for a full mapping of the PSD. Another important complementary diagnostic method is the at-wavelength reflectometry, by which the sensitivity for the frequency of the (micro-) roughness is selected by the x-ray wavelength of the radiation.

It is shown that reflectometry is a multi-dimensional experimental challenge: the photon energy and the incidence angle can be selected to give insight into the properties of the substrate and its contamination. The amount of specular radiation and the scattered light distribution give information on the optical quality of the surface (up to sandpaper quality).

Reflectometry is a powerful and indispensable tool for a quality check of optical elements. It delivers information on the slope and roughness of the surface and the optical properties - at the design wavelength. Sample alignment, however, is a serious issue - a source of misinterpreting the measured data.

Scattering of rough surfaces requires a complex mathematical treatment on a statistical basis.

At the BESSY-II optics beamline we have measured photon reflectivity and photo yield (electrons/photon) data for Al, Cu and SS samples to be used as materials for accelerator tubes of high energy machines. Agreement of the measured data with expectation and with modeling data is good.

# Acknowledgement

We thank M. Palmer, who stimulated this work in the framework of ILC-DR collboration. Takuya Ishibashi of KEK Accelerator Laboratory, Japan, during the BESSY-II beam time period on leave at LEPP, Cornell University, and Sara Casalbuoni of ANKA storage ring at KIT Karlsruhe, Germany are to be thanked for providing samples and for assistance during the measurements, and Laura Boon and V. Baglin provided samples of APS and LHC respectively. Support of the colleges of "Institute for Nanometre Optics and Technology", of Silvio Künstner for the measurements of the high-quality Si optics samples, and of Frank Siewert for providing these samples as well as for fruitful discussion about the slope error and roughness formalism is gratefully acknowledged.